\documentclass{osa-article}

\journal{oe}
\usepackage{multirow}

\articletype{Research Article}

\begin{document}

\title{Two-beam Coupling in the Production of Quantum Correlated Images by Four-wave Mixing}

\author{Meng-Chang Wu,\authormark{1,2} Nicholas R. Brewer,\authormark{1} Rory W. Speirs,\authormark{1} Kevin M. Jones,\authormark{3} Paul D. Lett\authormark{1,2,4,*}}

\address{\authormark{1}Joint Quantum Institute, National Institute of Standards and Technology and the University of Maryland, College Park, Maryland 20742, USA\\
\authormark{2}Chemical Physics Program, University of Maryland, College Park, Maryland 20742, USA\\
\authormark{3}Department of Physics, Williams College, Williamstown, Massachusetts 01267, USA\\
\authormark{4}Quantum Measurement Division, National Institute of Standards and Technology, Gaithersburg, Maryland 20899, USA}

\email{\authormark{*}paul.lett@nist.gov} 



\begin{abstract}
We investigate the effect of 2-beam coupling in different imaging geometries in generating intensity-difference squeezing from four-wave mixing (4WM) in Rb atomic vapors. A recently-introduced dual-seeding technique can cancel out the classical noise in a seeded four-wave mixing process. This dual-seeding technique, however, can introduce new complications that involve 2-beam coupling between different seeded spatial modes in the atomic vapor and can ruin squeezing at frequencies on the order of the atomic linewidth and below. This complicates some forms of quantum imaging using these systems. Here we show that seeding the 4WM process with skew rays can eliminate the excess noise caused by 2-beam coupling. To avoid 2-beam coupling in bright, seeded images, it is important to re-image the object in the gain medium, instead of focussing through it.
\end{abstract}

\section{Introduction}


The generation of two-mode or ``twin beam'' squeezing and entanglement from four-wave mixing (4WM) in atomic vapors has led to a number of interesting scientific results~\cite{Marino2009, Boyer2008, Hudelist2014, Anderson2017, Lawrie2019}, but also has promise for potential applications outside of the laboratory, such as in high-sensitivity detection and imaging~\cite{Lawrie2019}.  Sub-shot-noise correlations or entanglement between the beams generated in this system can potentially be used in random number generation and distribution, continuous-variable quantum communications, and metrological advances~\cite{Michel2019, Vernon2019, Marino2011}.  


In particular, sub-shot-noise differences of images could be produced by integrating pulses of light on a camera.  Such images at the present are limited by the accumulation of scattered light and low-frequency technical noise on the light.  Squeezing needs to be maintained over the whole optical spectrum of the pulse.  Experiments have shown that for typical pump laser intensities and detunings the squeezing bandwidths in these systems typically extend up to about 20 MHz, the cut-off depending on the intensity of the pump laser~\cite{Liu2011}.  The low-frequency cut-off of the squeezing has been in the kHz range, but recent advances have succeeded in obtaining low-frequency squeezing to about 10 Hz using a dual-seeding technique (balancing two seeded beams and their conjugates on the detectors)~\cite{Wu2019}.  This has now opened up the possibility of extending this technology to imaging applications.  Here we investigate a 2-beam coupling process that can degrade the squeezing that is obtained in these systems at low frequencies (of the order of the atomic transition linewidth).  

The 4WM process that generates the twin beams consists of a relatively intense pump beam and a weak probe beam crossing in a vapor cell.  Quantum correlations are generated between the probe beam, which is amplified by the 4WM process, and the conjugate beam that is generated in the parametric gain process.  The dual-seeding technique that can remove low-frequency noise in the intensity-difference squeezing spectrum generated from 4WM when the twin beams are detected in their entirety brings along with it new complications that involve 2-beam coupling between the beams if the beams are detected in part, say as several pixels in probe and conjugate images.  The 2-beam coupling effect generates large amounts of excess noise as a result of fluctuations in the atomic polarization seen by one beam that are driven by a second beam, and has been previously observed in similar situations~\cite{Agarwal1988, Kauranen1994} between beams that are degenerate in frequency, are tuned near a transition resonance, and that cross in an atomic vapor.  In the present case if the seed beams for the 4WM process (which are amplified in the pumped Rb vapor) cross in the vapor cell, the quantum correlations that the 4WM gain produces will be compromised.  The vacuum sidebands of one beam can interact with another beam in the atomic vapor and, exchanging energy, amplify the intensity noise through 2-beam coupling.  This process becomes significant when the intensity of the seeding beam is larger than the saturation intensity of the atomic transition~\cite{Kauranen1994}.

We show that similar 2-beam coupling can also be present in the dual-seeding situation discussed here.  The addition of the pump and the 4WM interaction that amplifies the fields of the seed beams enhances this coupling, to the detriment of the observed squeezing at frequencies below the atomic transition linewidth.  We find that the issue can be avoided either with sufficiently low power (in both the probe input seed and the output amplified probe - as was the situation in Ref.~\cite{Wu2019} where this complication did not appear) or by seeding with beams that do not directly intersect in the gain medium.

This problem is especially relevant to the case of generating quantum-correlated images.  To have the input probe (image) completely within the highest gain region, we can focus the probe into the center of the pump beam in the Rb cell in Fig.~\ref{fig:fig42} (a), placing the Fourier plane of the optical system in the gain region.  Under these circumstances the information in the image can be distributed into all of the pixels in the Fourier plane and this will cause ``cross-talk" between the pixels of the image in the 4WM process.  We can eliminate this problem by imaging the probe seed into the 4WM gain region in Fig.~\ref{fig:fig42} (b).  In that case the light in each pixel of such an image travels parallel to the light in the other pixels and does not overlap and exchange energy with the other pixels by interacting with the same atoms.  We present experiments demonstrating the 2-beam coupling as well as the imaging solution to the problem.

\section{Experiments}

\begin{figure}
\centering
\includegraphics[scale=0.65]{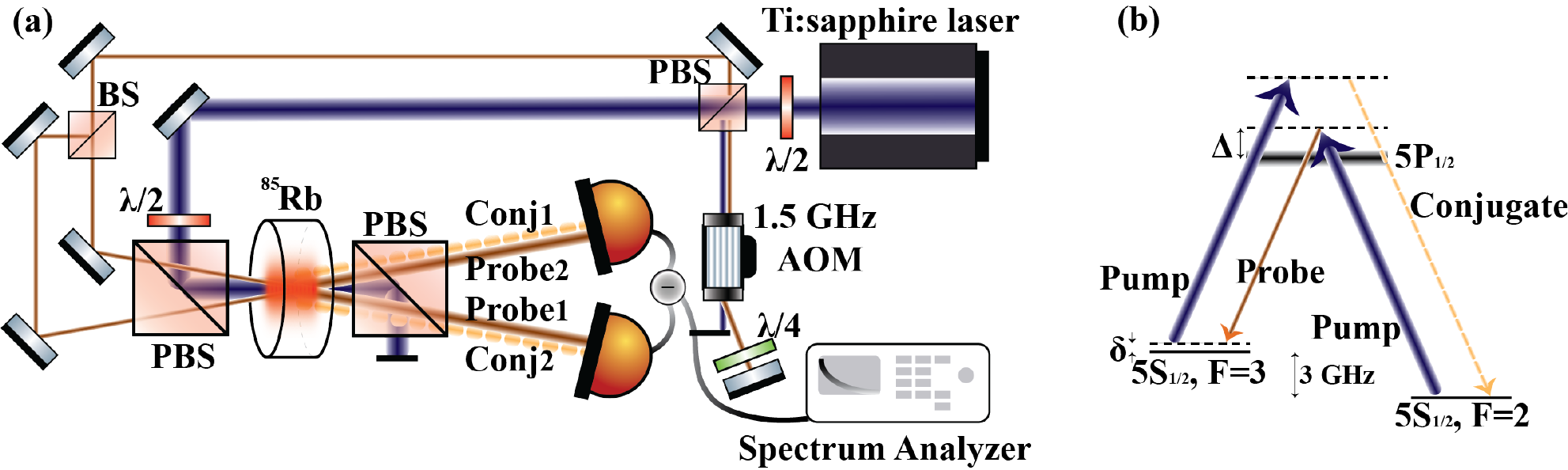}
\caption{(a) Experimental setup for studying 2-beam coupling.  BS indicates a non-polarizing beamsplitter, PBS a polarizing beamsplitter.  $\lambda$/2 and $\lambda$/4 are half-wave and quarter-wave plates.  AOM is a 1.5-GHz acousto-optic modulator.  (b) Energy level diagram for the non-degenerate 4WM process in $^{85}$Rb.  The optical wavelength is $\lambda$ = 795  nm.  The broadened upper level indicates the Doppler broadening of the transition.  Typical detunings are $\Delta$ = 800 MHz to 1.3 GHz and  $\delta$ = -2 MHz.}
\label{fig:fig41}
\end{figure}

The experimental setup that we use to demonstrate the 2-beam coupling in the squeezed light generation (Fig.~\ref{fig:fig41} (a)) is almost identical to the setup used to demonstrate low-frequency squeezed light~\cite{Wu2019}.  We use a Ti:sapphire laser as the basis for these experiments because of its superior noise properties in these experiments in comparison to most diode lasers.  The light from the Ti:sapphire laser is sent through an optical fiber to create a pump beam of $\approx1.5$ mm $1/e^2$ diameter and \hbox{650 mW} of power.  This light is detuned in the range of $\Delta=800$ MHz to \hbox{1.3 GHz} to the blue of the S$_{1/2}$ (F=2) $\rightarrow$P$_{1/2}$ (F=3) transition in $^{85}$Rb.  The $^{85}$Rb cell is \hbox{1.2 cm} long and is heated to a temperature of $\approx125$ $^\circ$C.  A pair of seed beams at the probe frequency can be introduced on either side of the pump, at a small angle ($\approx0.3$ degrees to 0.5 degrees) to the pump beam.  The probe seeds are 0.55 mm $1/e^2$ diameter beams derived directly from the Ti:sapphire laser by double-passing $\approx90$ mW of this light through a 1.5 GHz acousto-optic modulator, resulting in a 2-photon detuning of $\delta=-2$ MHz for the 4WM process.

\begin{figure}
\centering
\includegraphics[scale=1]{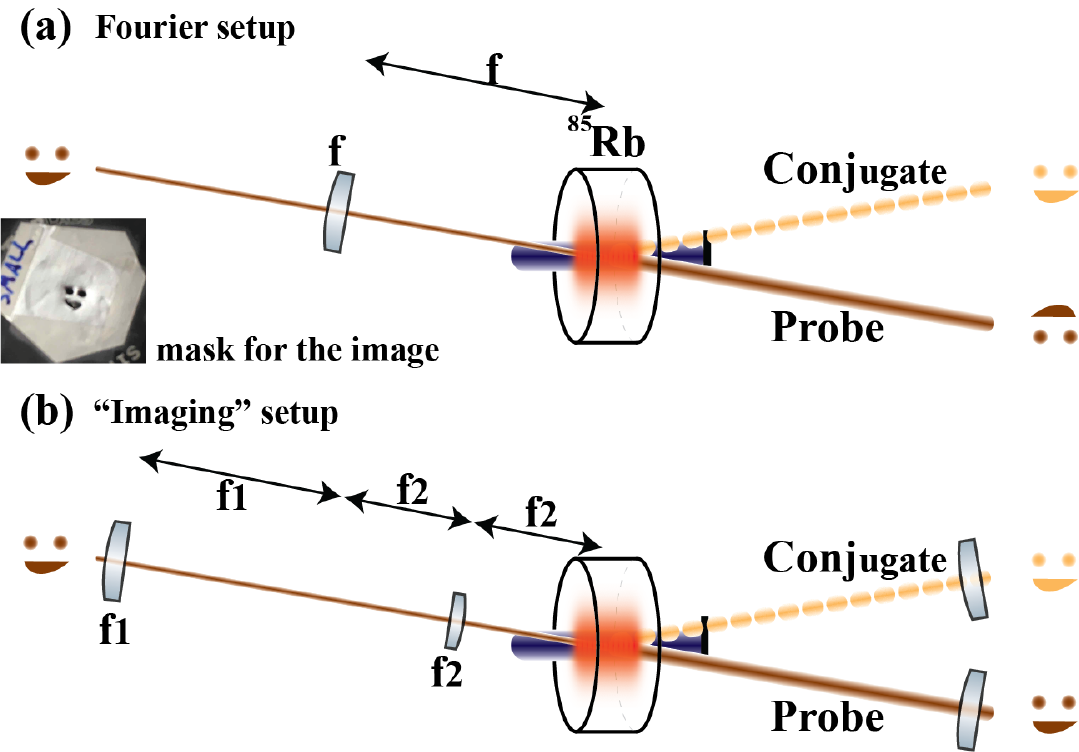}
\caption{Quantum imaging setup.  The laser beam is sent through a ``smiley'' mask to generate the image seed.  (a) Fourier setup.  The Fourier plane of the image is focused into the 4WM gain region.  Without the focusing lens the probe does not fit into the 4WM gain region.  (b) ``Imaging'' setup.  A beam diameter reducing telescope creates a real image of the mask in the 4WM gain region.  Note that there is a Kerr lensing effect due to the 4WM gain medium that needs to be considered to understand the images that are observed in the far field.}
\label{fig:fig42}
\end{figure}

There are two experimental setups in Fig.~\ref{fig:fig42} for the quantum correlated imaging.  The image seed in real space is formed by sending a laser beam through a smiley mask.  The Fourier setup (Fig.~\ref{fig:fig42} (a)), is a typical setup for the 4WM, where a Gaussian probe seed beam is focused into the 4WM gain region.  A mask is inserted into the probe beam before the focussing lens.  With this setup for the imaging experiment, the information at the center of the Rb cell will be the probe image in spatial frequency (that is the Fourier transform of the image seed).  That means each component of the probe seed in the gain region contains all the information of the probe image with the same spatial frequency.  This causes 2-beam coupling between bright pixels when the mask is reimaged.  In Fig.~\ref{fig:fig42} (b), the imaging setup, the beam diameter reducing telescope demagnifies the probe beam to fit into the center of the 4WM gain region.  The most important function of the beam diameter reducing telescope is to form an image of the mask in real space in the 4WM gain region, which avoids the 2-beam coupling problem.  A 10 ns delay line is included in the conjugate image to compensate for the group velocity difference between the probe and conjugate beams.

\begin{figure}
\centering
\includegraphics[scale=0.65]{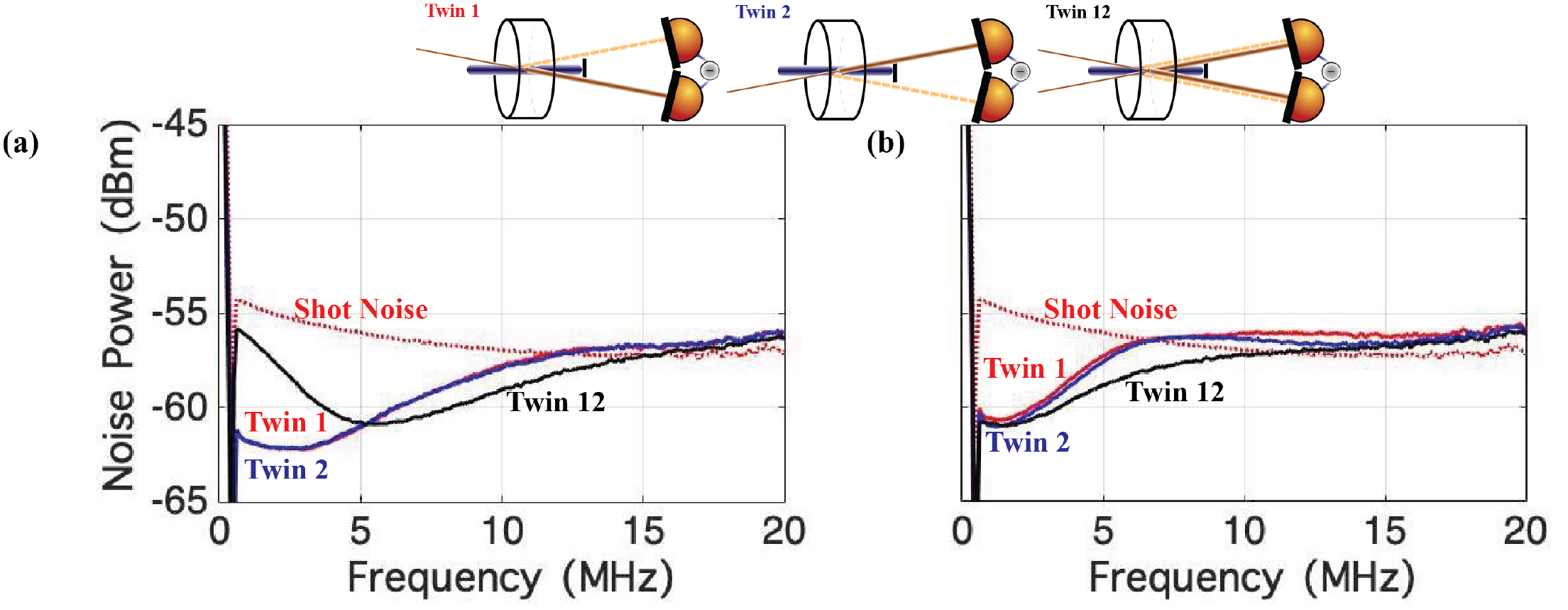}
\caption{Spectrum of the intensity-difference squeezing versus measurement frequency of (a) crossed rays and (b) skew rays.  Shown are shot noise (red dashed curve), spectra of two independent single seeded processes labeled ``Twin1'' (red curve) and ``Twin2'' (blue curve), and dual seeded 4WM labeled ``Twin12'' (black curve).  Crossed rays: both of these seeds are present at half the power of the single seeds and crossed at the center of the pump beam in the 4WM gain region.  Skew rays: two probe beams are offset from the center of the pump beam in opposite directions.  The total output optical power for each trace is \hbox{800 $\mu$W}.  The 4WM gain $\approx10$ in each case (pump power = \hbox{680 mW}, cell temperature = 122 $^\circ$C, $\Delta =800$ MHz, $\delta = -2$ MHz).  There is a delay line for the conjugate beams in each case in order to compensate for the group delay difference introduced by the Rb cell.  The resolution bandwidth (RBW) is 300 kHz and the video bandwidth (VBW) is 100 Hz for these measurements.  The detector bandwidth is 45 MHz.  The electronic noise is subtracted from these traces.}
\label{fig:fig43}
\end{figure}

\section{Results}

While the dual-seed technique clearly improves the squeezing at low frequencies, it also introduces new problems, where the squeezing is now noticeably reduced at frequencies below the atomic transition linewidth, shown in Fig.~\ref{fig:fig43} (a).  A similar effect has been observed in the past: introducing two bright beams to seed a 4WM process with a single pump can couple the beams and affect the observed squeezing.  This observation has led to the use of a completely separate pump beam to generate local oscillators (LOs) for homodyne detection in experiments where that was required~\cite{Boyer2008}, ensuring the independence of the two 4WM processes.

\begin{figure}
\centering
\includegraphics[scale=0.5]{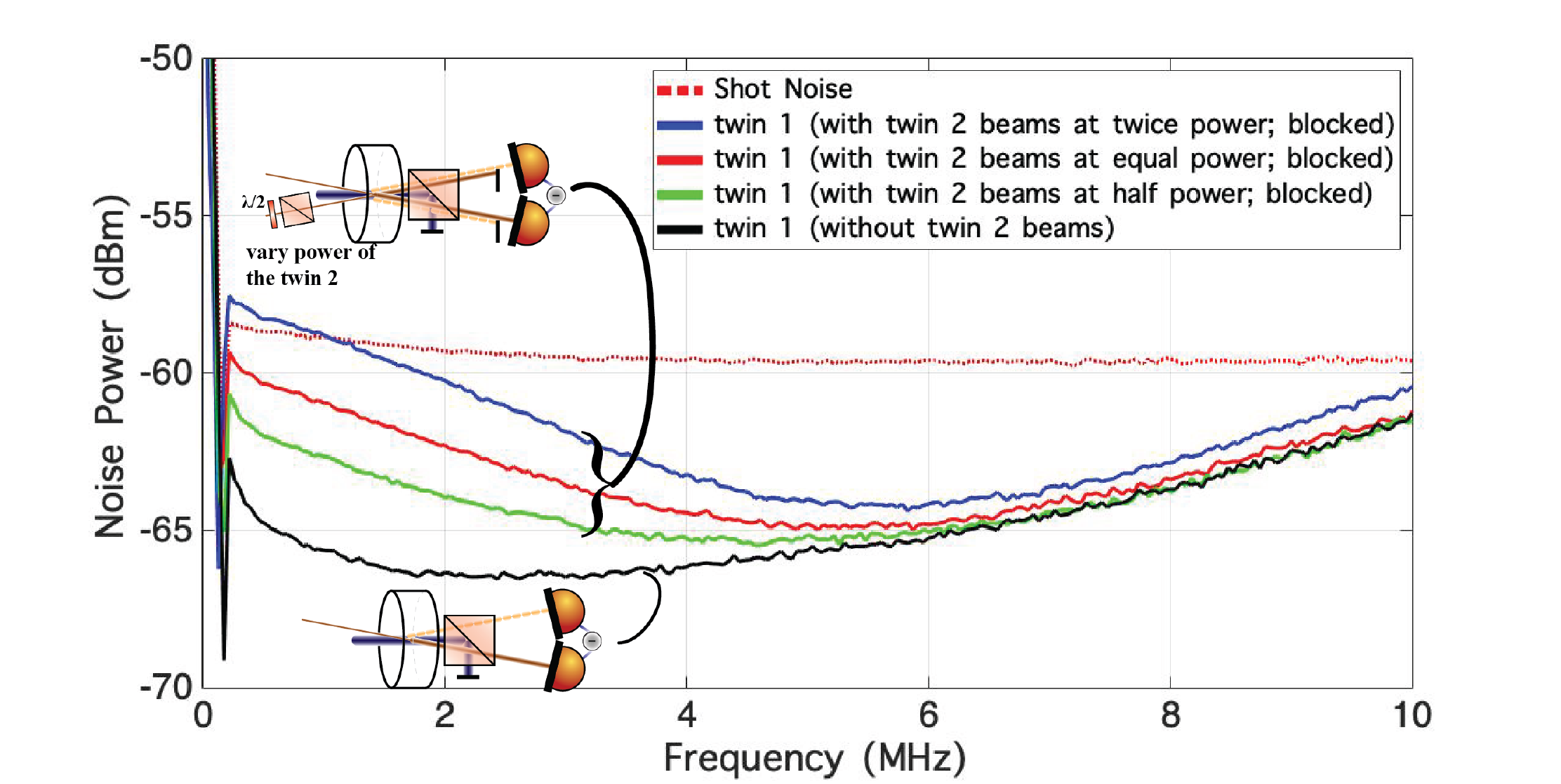}
\caption{2-beam coupling at 3 relative powers.  The plotted spectra are the shot noise measurement (red dashed curve); intensity-difference squeezing of one set of twin beams in the presence of a second set of twin beams that are blocked before the detectors.  The second beam pair has twice (blue curve), equal (red curve), and half power (green curve) of the measured beam pair, crossed in the 4WM gain region; intensity-difference squeezing of a single pair of twin beams with no second seed present (black curve).  The total output optical power for each trace is \hbox{800 $\mu$W}.  The 4WM gain $\approx7$ in each case (pump power = \hbox{750 mW}, cell temperature = \hbox{120 $^\circ$C}, $\Delta =800$ MHz, $\delta = -2$ MHz).  There is a delay line for the conjugate beams in each case.  The resolution bandwidth (RBW) is 100 kHz and the video bandwidth (VBW) is \hbox{100 Hz} for these measurements.  The detector bandwidth is \hbox{45 MHz}.  The electronic noise is subtracted from these traces.}
\label{fig:fig44}
\end{figure}

The dual-seeding technique can be seen to markedly reduce the squeezing at frequencies below about 5 MHz in Fig.~\ref{fig:fig43} (a).  It seems that there is a coupling between the seeded beams that degrades the squeezing in this regime.  We have investigated this effect as a function of the relative beam powers, as shown in Fig.~\ref{fig:fig44}.  Here we fix the power of one of the seed beams and vary the power of the second input seed, showing that the coupling between the beams increases with intensity.

\begin{figure}[!t]
\centering
\includegraphics[scale=0.6]{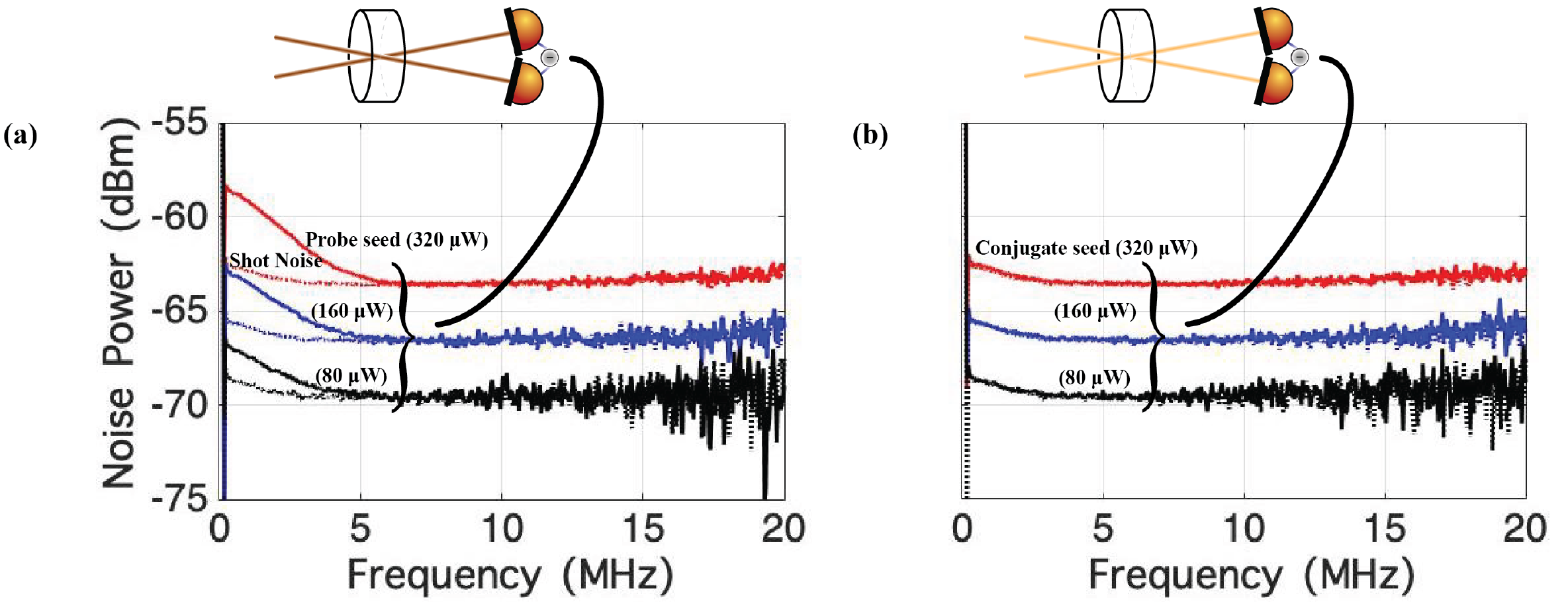}
\caption{Noise spectra of the intensity-difference of (a) two probe seed beams (detuned $\approx800$ MHz to the blue of the S$_{1/2}$ (F=3) $\rightarrow$P$_{1/2}$ (F=3) transition in $^{85}$Rb), (b) two seed beams at the conjugate frequency ($\approx4$ GHz to the blue of the S$_{1/2}$ (F=2) $\rightarrow$P$_{1/2}$ (F=3) transition in $^{85}$Rb) crossing in the Rb vapor cell and the shot noise (dashed curves) in free space.  Without the Rb vapor present the technical noise on the beams subtracts and gives a measure of the shot noise for the incident power.  With the Rb vapor present the 2-beam coupling causes the probe seed beams to exchange energy and increases the noise power in the intensity-difference at low frequencies.  The resolution bandwidth (RBW) is 100 kHz and the video bandwidth (VBW) is 100 Hz for these measurements.  The detector bandwidth is 45 MHz.  The electronic noise is subtracted from these traces.}
\label{fig:fig45}
\end{figure}

The 2-beam coupling observed here is similar to that reported by Kauranen \textit{et al.}~\cite{Kauranen1994}.  In those experiments two beams of the same frequency, tuned near an atomic resonance, cross in a vapor cell and exchange energy, resulting in enhanced fluctuations in the output intensity of each beam.  The excess noise in those experiments was explained in terms of a 4WM interaction in a 2-level atom model~\cite{Agarwal1988}.  The model allowed for two noise mechanisms.  The first is amplification of vacuum fluctuations in the frequency sidebands of one beam, due to a 4WM interaction with the other beam (a stronger ``pump'' beam in their case).  In this case the amplified sidebands beating against the carrier of the injected beam produces a beam that is much noisier than the shot-noise-limited input at low frequencies.  A second mechanism proposed by these authors is the spontaneous scattering of light resulting from quantum fluctuations of the atomic medium that are induced by the other beam.  This mechanism, basically the absorption by the atoms of light from one beam that results in resonance fluorescence being re-emitted into the spatial modes of the other beam.  This light is also further amplified by a 4WM interaction.  The first mechanism is only expected to occur when the light intensity is comparable to the saturation intensity of the transition and the latter mechanism was determined to be the dominant one in~\cite{Kauranen1994}.  In the present case the intensity of each of the two degenerate probe beams is generally well below the saturation intensity of the Rb atomic transition, discriminating against the first mechanism.  The fact that the coupling is restricted to frequencies at or below the natural linewidth of the atomic transition (the linewidth of the P$_{1/2}$ state in Rb is $\approx$ 5.75 MHz) as observed in Fig.~\ref{fig:fig45} (a) is consistent with this interpretation.


The 2-beam coupling mechanism occurs with or without the pump beam present and is a 4WM coupling between the degenerate probe beams, mediated by the atomic vapor.  This is illustrated in Fig.~\ref{fig:fig45} (a) where the two probe seed beams, which are 800 MHz blue of the transition from the higher ground state ($F_g = 3$) to the excited state, are intersected in free space (to determine the shot noise level) and in a Rb cell, with no pump present, at a series of increasing powers.  The 2-beam coupling effect is a nonlinear process that depends on the intensity of the seed beams.  The excess noise appears over the same frequency range as in Fig.~\ref{fig:fig44} where the pump is present, and the amount of excess noise increases with power.  This behavior is again consistent with the 2-beam coupling explanation given above.  However, the 2-beam coupling effect does not happen when the degenerate beams are far off resonance.  For example, in Fig.~\ref{fig:fig45} (b) we do not see the excess noise when the seed beams are at the conjugate frequency, which is $\approx$ 4 GHz blue of the transition from the lower ground state ($F_g = 2$) to the excited state.


The work in~\cite{Kauranen1994} was carried out before the current generation of experiments in which strong squeezing was observed using atomic systems such as those used here, and it is interesting to note that these authors state, ``We believe that this mechanism can be important in preventing the reduction of noise below the quantum-noise limit in experiments that utilize atomic vapors.''  We can see here that these concerns were perhaps more sweeping than justified, at least for non-degenerate 4WM processes such as that employed in the present work, where strong squeezing can be obtained.  It is, however, clearly a prescient concern in terms of the coupling of 4WM processes in a multi-spatial-mode (imaging) system.  It is also unclear whether the 4WM process in which frequency-degenerate twin beams are generated (essentially swapping the roles of the pump and the probe and conjugate frequencies used here) would also suffer from this 2-beam coupling noise~\cite{Jia2017}.

The 2-beam coupling was not observed, for instance, in Ref.~\cite{Wu2019}, where very low seed intensities were used, so that the coupling, even at the output, was never very large.  The 2-beam coupling effect only happens when the two seed beams interact with the same atoms.  An obvious way to keep the two probe beams from interacting is to send the two probe beams through the gain medium as skew rays, without intersecting each other.  In that way independent groups of atoms contribute to the two 4WM processes and this coupling mechanism should be eliminated.  Figure~\ref{fig:fig43} shows the results of this comparison.  Figure~\ref{fig:fig43} (a) shows the spectra of the two independent 4WM processes (red and blue curves), as well as the simultaneous dual-seed process (black curve) with intersecting beams that displays a large degradation in the squeezing at frequencies below about 5 MHz.  Figure~\ref{fig:fig43} (b) shows spectra taken under very similar conditions except that the beams have been moved so that the seed beams do not intersect.  In this case the simultaneous dual-seeded process shows improved squeezing at frequencies below \hbox{5 MHz} compared to both single-seeded and dual-intersecting-seeded 4WM processes. 

The gain bandwidth of the 4WM is partially determined by the pump power.  Figure~\ref{fig:fig43} (b) shows that the squeezing with the dual-seeding technique only extends to \hbox{10 MHz}, which is less than the squeezing bandwidth of 15 MHz in Fig.~\ref{fig:fig43} (a).  In the dual-seeded 4WM with skew rays the probe seed beams are not at the center of the pump beam in the 4WM gain region, therefore, the 4WM bandwidth is smaller due to the weaker pump intensity in the 4WM process.  At the measurement frequency of 6 MHz, where the 2-beam coupling effect is not present, the skew rays scheme in Fig.~\ref{fig:fig43} (b) has 2 dB less squeezing than the squeezing in Fig.~\ref{fig:fig43} (a) with the probe beams at the center of the pump.  The problems of smaller 4WM bandwidth and less squeezing can be avoided by having more pump power or having flat-top beam-shaping optics for the pump.

The low-frequency 2-beam coupling mechanism discussed above can be avoided for the simple case of two ``beams'' that need to be independent.  This could include, for instance, the generation of local oscillator beams for homodyne detection, which need to be independent of the entangled signal beams used in many experiments.  While the beam distortions from the Kerr lensing due to the pump beams could be difficult to match if the signal and local oscillator beams pass through different parts of the pump region, this could perhaps be fixed with a flat-top intensity profile for the pump.  Unfortunately, this 2-beam coupling has significant negative implications for imaging applications, as it is a mechanism for cross-talk between the pixels in an imaging field.  If a masked or greyscale image is used to seed a 4WM process (such as in~\cite{Boyer2008, Clark2012, Lawrie2013}), the seed image is typically focused into the gain medium and re-imaged after emerging, as in \hbox{Fig.~\ref{fig:fig42} (a)}.  If two distinct spatial modes (pixels) that are part of an input image are bright they will then necessarily overlap within the gain medium and experience coupling to the extent that the spatial Fourier transform overlaps them.  The coupling can be avoided to some degree by imaging the seed pattern into the gain medium, shown in \hbox{Fig.~\ref{fig:fig42} (b)}, so that the spatial modes largely do not cross and couple in the gain medium.

\begin{figure}[!t]
\centering
\includegraphics[scale=1]{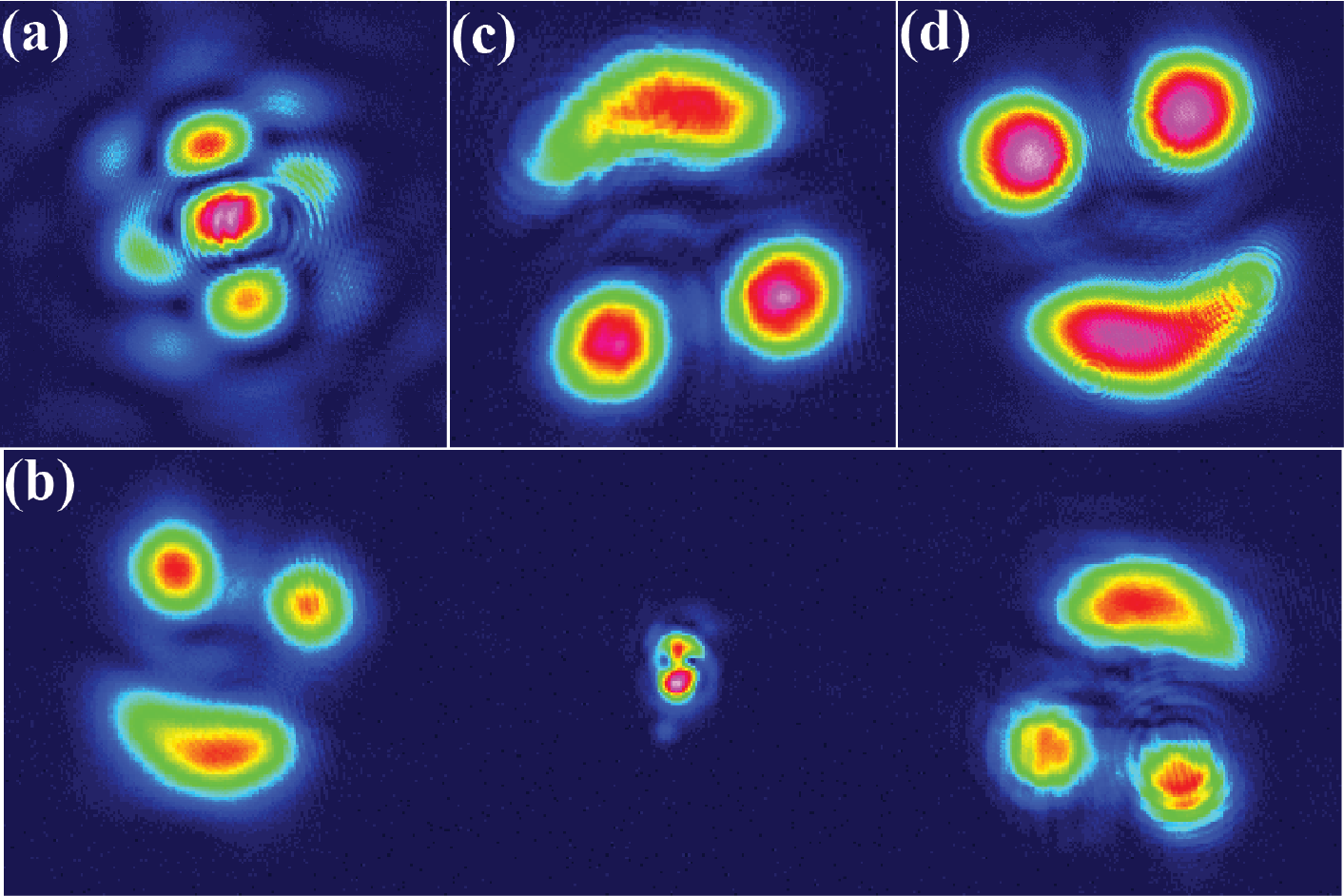}
\caption{Images in the Fourier setup (Fig.~\ref{fig:fig42} (a)).  (a) The transform of the smiley face in k-space at the 4WM gain region.  The pump beam size is \hbox{1.55 mm} $1/e^2$ diameter and the main area of the smiley face in k-space is half of the diameter of the pump.  (b) Image of the conjugate (left) and the probe (right) in the far field.  The middle spot is the depolarized pump light.  (c) The re-imaged probe and (d) the re-imaged conjugate.  There is a polarizing beamsplitter before the vapor cell.  With a wave plate we can redirect the image to take a picture at the equivalent distance to the cell position.}
\label{fig:fig48}
\end{figure}

\begin{figure}[!t]
\centering
\includegraphics[scale=0.6]{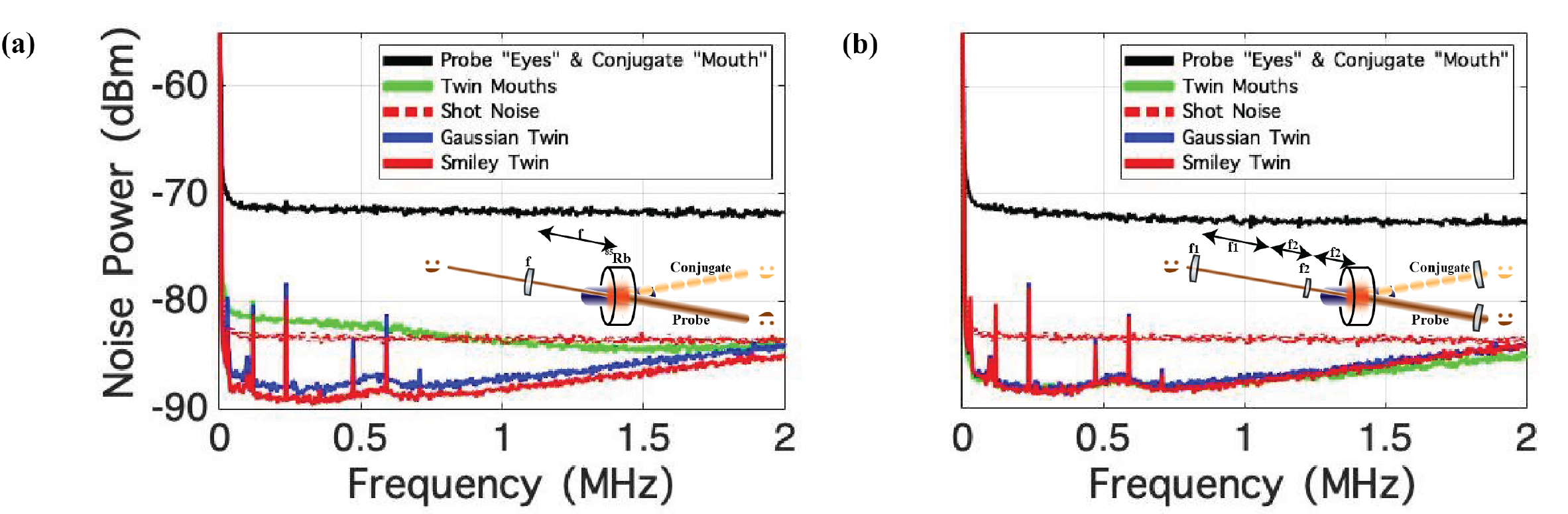}
\caption{Squeezing spectrum obtained using (a) the Fourier setup (Fig.~\ref{fig:fig42} (a)) and (b) the imaging setup (Fig.~\ref{fig:fig42} (b)).  The pump beam is at a one-photon detuning of 1.3 GHz, the gain is $\approx6$, and no delay lines are inserted in any of the beams for these measurements.  The two-photon detuning is kept at \hbox{-2 MHz} and the pump power is 650 mW.  The total output power for each trace is $\approx500$ $\mu W$.  The traces are the thermal noise of the ``eyes'' in the probe and the ``mouth'' in the conjugate (black curve); intensity-difference squeezing of twin mouths (green curve); the shot noise measurement (red dashed curve); intensity-difference squeezing of Gaussian probe beam (blue curve); intensity-difference squeezing of the whole smiley face (red curve).  The size of smiley face is smaller than the Gaussian beam in (a).  The resolution bandwidth (RBW) is 1 kHz and the video bandwidth (VBW) is \hbox{100 Hz} for these measurements.  The detector bandwidth is 4 MHz.  The electronic noise is about 20 dB below the shot noise level and is not subtracted from these traces.}
\label{fig:fig49}
\end{figure}

The ``whole beam'' measurements in references~\cite{Boyer2008, Clark2012, Lawrie2013} did not show this effect -- that is, if you seed multiple spatial modes (or completely separate beams) but integrate over these beams, the loss of correlations between individual pixels is not apparent.  In that case, while 2-beam coupling causes, say, two beams or spatial modes at the probe frequency to exchange energy, the correlations are preserved if we integrate over both beams on a single detector.  The correlated photons are not lost, they are just moved into another spatial mode, and in that case this mode is also detected.  In the present case, however, we integrate over probe 1 and \hbox{conjugate 2} on a single detector, and now when energy is exchanged between the two probe beams the correlations are rearranged between the two detectors and the squeezing is reduced.


Figure~\ref{fig:fig48} shows the k-space image in the 4WM gain region, quantum correlated images in the far field, and the re-imaged outputs obtained using the Fourier setup as shown in Fig.~\ref{fig:fig42} (a).  In Fig.~\ref{fig:fig48} (b) the conjugate image is rotated by 180$^\circ$ because of the momentum conservation in k-space.  In the 4WM process the probe frequency is much closer to the atomic line than the conjugate.  Given this smaller detuning of the probe compared to the conjugate, we expect to see stronger nonlinear effects on the probe beam.  The probe image experiences a self-focusing effect that makes the focus position of the probe and conjugate slightly different in Fig.~\ref{fig:fig48} (b).  To check the 2-beam coupling effect we measure the correlation between the twin images and compare it with the squeezing spectrum of the individual Gaussian twin beams.  In Fig.~\ref{fig:fig49} (a) the squeezing spectrum of the whole smiley face is similar to that of the Gaussian twin beams, as expected.  The low frequency technical noise in the shot noise measurement is present in the squeezing spectrum of the smiley twin (red) and the Gaussian twin (blue).  Table~\ref{tab:squeezingtable} shows the squeezing of the two configurations at a measurement frequency of 1 MHz.  The main difference between two configurations is that the Fourier configuration loses squeezing between individual features (such as the eyes or mouths) of the probe and the conjugate images at low frequencies due to the 2-beam coupling effect.  The probe or the conjugate beam itself is a displaced thermal beam.  The intensity difference noise of different parts of the probe and the conjugate images will result in thermal noise.  The comparison in Table~\ref{tab:squeezingtable} demonstrates that the 2-beam coupling effect causes the energy exchange between pixels of an image and still maintains the quantum correlations between the whole images.


\begin{figure}[!t]
\centering
\includegraphics[scale=1]{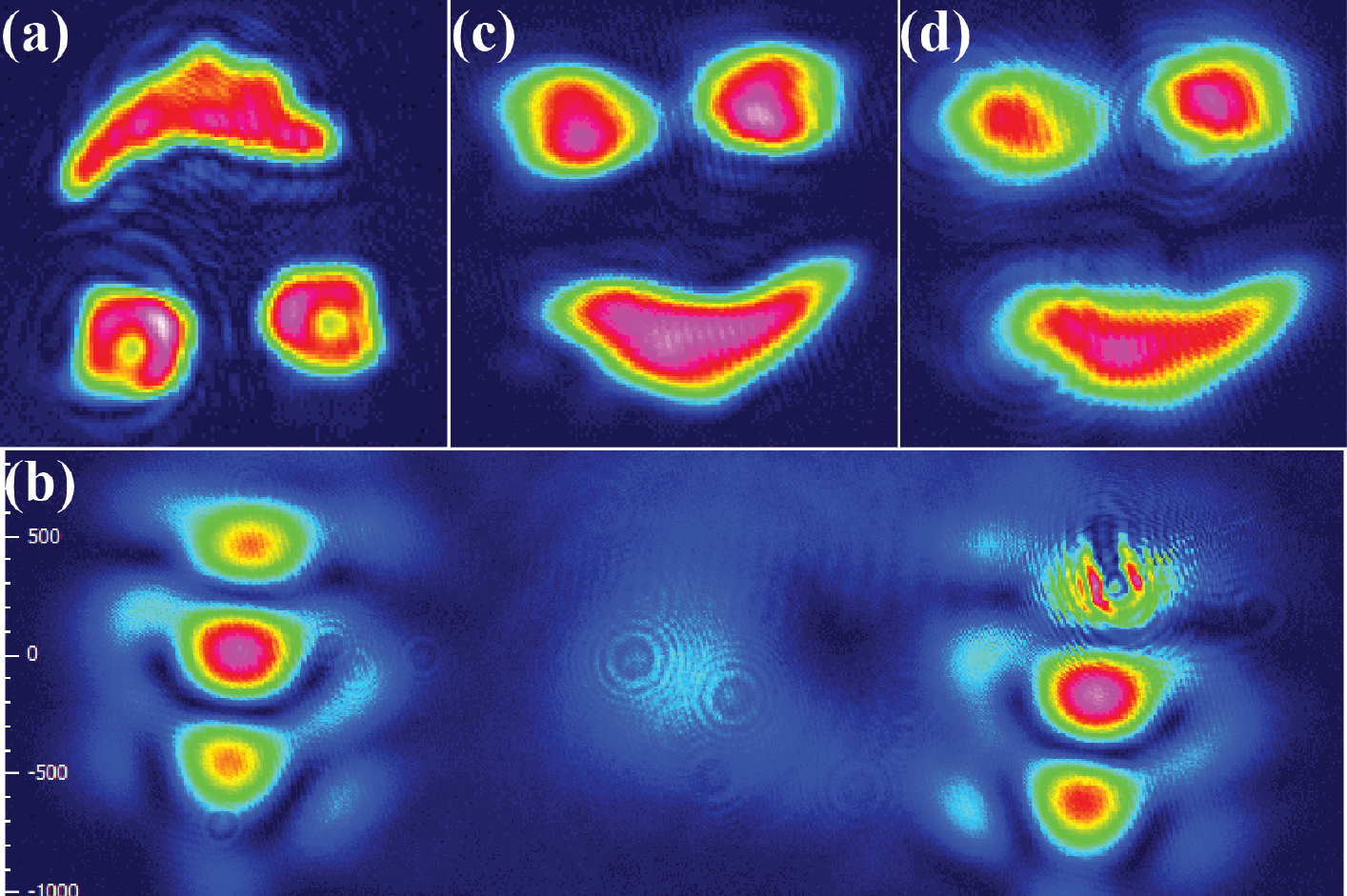}
\caption{Images obtained from the imaging setup of Fig.~\ref{fig:fig42} (b).  (a) The smiley face is mapped to the 4WM gain region with a beam diameter reducing telescope.  The pump beam size is 1.55 mm $1/e^2$ diameter and the size of the smiley face is around one third of the diameter of the pump.  (b) Image of the conjugate (left) and the probe (right) in k-space in the far field.  The middle spot is the depolarized pump light.  (c) The re-imaged probe and (d) the re-imaged conjugate.  There is a defect in the camera on the upper right of the image (b).}
\label{fig:fig411}
\end{figure}


Figure~\ref{fig:fig411} shows images obtained using the imaging setup of Fig.~\ref{fig:fig42} (b).  The probe image in Fig.~\ref{fig:fig411} (a) is taken at the position of the 4WM gain.  In this case, each pixel of the image does not overlap with the others in the gain medium and this eliminates cross talk between the pixels.  Because of the Kerr lensing in the vapor cell in this setup, the twin images in the far field are in the k-space shown in Fig.~\ref{fig:fig411} (b) and this makes it difficult to separate the twin images when high spatial frequencies are present.  Choosing a larger phase-matching angle can help avoid this problem.  The only disadvantage of the large phase-matching angle is the smaller 4WM gain bandwidth.  The correlations between the whole images obtained using the imaging setup shown in Fig.~\ref{fig:fig49} (b) give the same result as the one using the Fourier setup, that is, their squeezing spectra are similar to the squeezing spectrum of the Gaussian twin beams.  The imaging setup also avoids the exchange of energy between the pixels in the probe image caused by the 2-beam coupling effect.


\begin{table}[]
\begin{tabular}{cccccccccc}
(a)                                         &                           & \multicolumn{3}{c}{Conjugate}                                                   & (b)                                         &                           & \multicolumn{3}{c}{Conjugate}                                                 \\ \cline{2-5} \cline{7-10} 
\multicolumn{1}{c|}{}                       & \multicolumn{1}{c|}{(dB)} & \multicolumn{1}{c|}{ \raisebox{-.3\height}{\includegraphics[width=0.45 in]{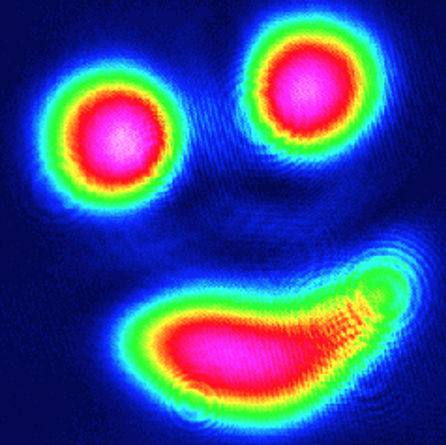}} }     & \multicolumn{1}{c|}{ \raisebox{-.3\height}{\includegraphics[width=0.450 in]{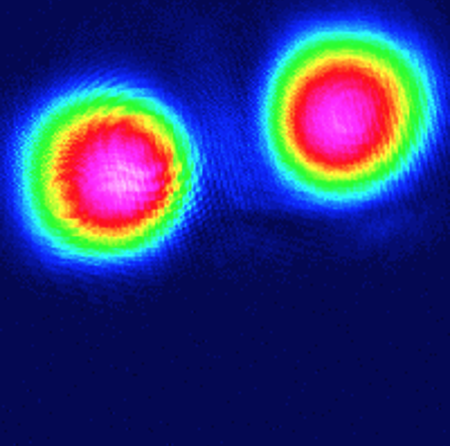}} }     & \multicolumn{1}{c|}{ \raisebox{-.3\height}{\includegraphics[width=0.450 in]{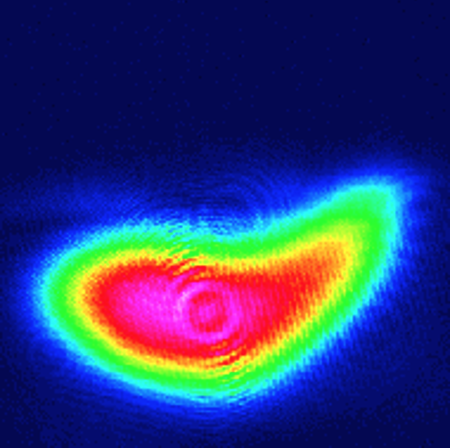}} }   & \multicolumn{1}{c|}{}                       & \multicolumn{1}{c|}{(dB)} & \multicolumn{1}{c|}{ \raisebox{-.3\height}{\includegraphics[width=0.45 in]{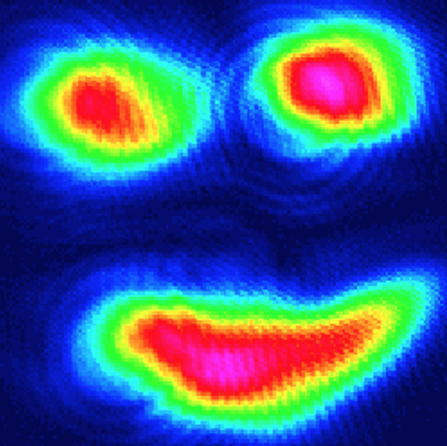}} }   & \multicolumn{1}{c|}{ \raisebox{-.3\height}{\includegraphics[width=0.45 in]{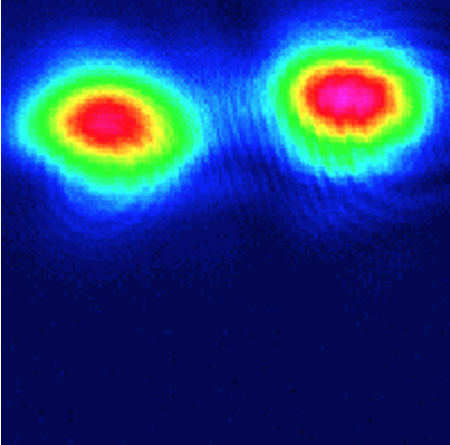}} }     & \multicolumn{1}{c|}{ \raisebox{-.3\height}{\includegraphics[width=0.45 in]{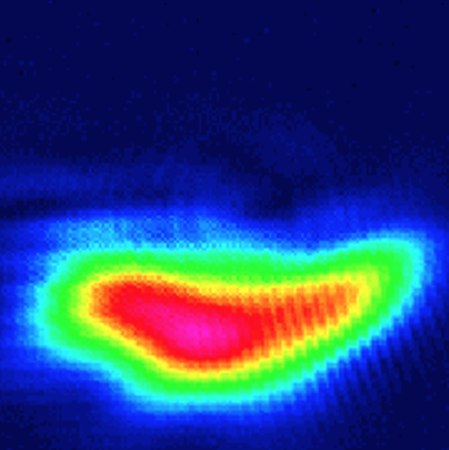}} }   \\ \cline{2-5} \cline{7-10} 
\multicolumn{1}{c|}{\multirow{3}{*}{ \rotatebox[origin=c]{90}{Probe} }} & \multicolumn{1}{c|}{ \raisebox{-.3\height}{\includegraphics[width=0.45 in]{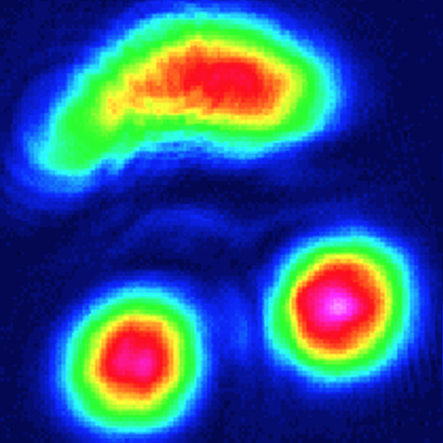}} }     & \multicolumn{1}{c|}{-4.5} & \multicolumn{1}{c|}{x}    & \multicolumn{1}{c|}{x}  & \multicolumn{1}{c|}{\multirow{3}{*}{ \rotatebox[origin=c]{90}{Probe} }} & \multicolumn{1}{c|}{ \raisebox{-.3\height}{\includegraphics[width=0.45 in]{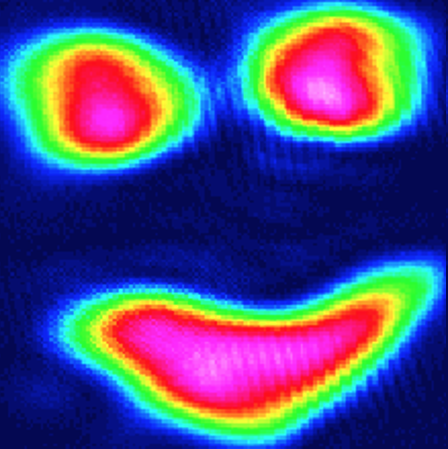}} }     & \multicolumn{1}{c|}{-4} & \multicolumn{1}{c|}{x}    & \multicolumn{1}{c|}{x}  \\ \cline{2-5} \cline{7-10} 
\multicolumn{1}{c|}{}                       & \multicolumn{1}{c|}{ \raisebox{-.3\height}{\includegraphics[width=0.44 in]{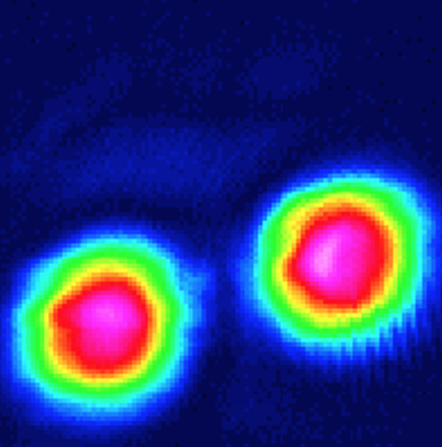}} }     & \multicolumn{1}{c|}{x}    & \multicolumn{1}{c|}{-0.5} & \multicolumn{1}{c|}{12} & \multicolumn{1}{c|}{}                       & \multicolumn{1}{c|}{ \raisebox{-.3\height}{\includegraphics[width=0.45 in]{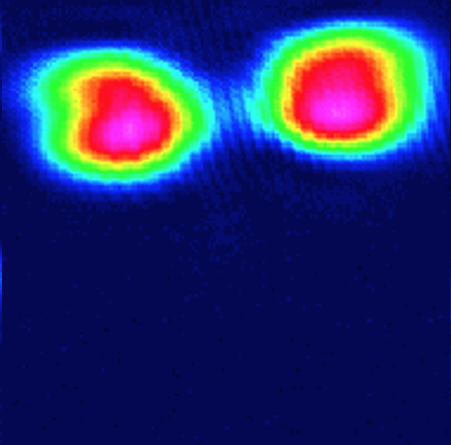}} }     & \multicolumn{1}{c|}{x}  & \multicolumn{1}{c|}{-3.5} & \multicolumn{1}{c|}{11} \\ \cline{2-5} \cline{7-10} 
\multicolumn{1}{c|}{}                       & \multicolumn{1}{c|}{ \raisebox{-.3\height}{\includegraphics[width=0.44 in]{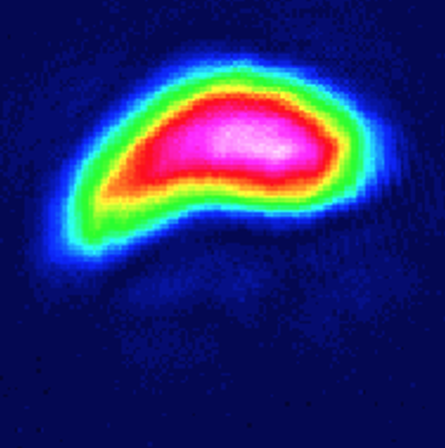}} }     & \multicolumn{1}{c|}{x}    & \multicolumn{1}{c|}{12}   & \multicolumn{1}{c|}{0}  & \multicolumn{1}{c|}{}                       & \multicolumn{1}{l|}{ \raisebox{-.3\height}{\includegraphics[width=0.45 in]{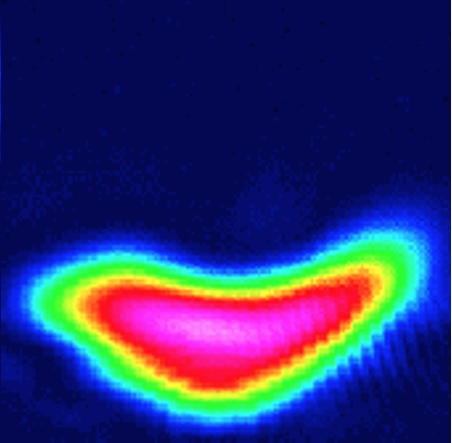}} }     & \multicolumn{1}{c|}{x}  & \multicolumn{1}{c|}{11}   & \multicolumn{1}{c|}{-4} \\ \cline{2-5} \cline{7-10} 
\end{tabular}
\caption{Squeezing measurements (in dB) between the probe and conjugate images at a frequency of 1 MHz for (a) the Fourier setup, and (b) the imaging setup.}
\label{tab:squeezingtable}
\end{table}

\section{Discussion}

The 2-beam coupling mechanism is discussed above starting from the coupling between two Gaussian beams with the same frequency that causes excess noise at measurement frequencies below the atomic transition linewidth.  We consider two Gaussian beams as two pixels of an image that can also have cross talk due to the 2-beam coupling effect if we do not map the image to the 4WM gain region properly. 

In the 4WM process we can focus a collimated Gaussian probe seed beam into the gain region as in the Fourier setup.  Then we get a pair of twin beams, the probe and conjugate beams, in real space.  We should see the 2-beam coupling effect degrades the squeezing if the twin beams are cut to half area from opposite sides of the beams in real space.  By contrast, in the imaging setup for the Gaussian twin images, instead, we expect to see that each pair of pixels of the images of the Gaussian twin beams are quantum correlated.  That is, we might be able to maintain a good amount of squeezing if we cut the Gaussian twin beams using the imaging setup with irises because we assume that the Gaussian beam is also a kind of a multi-spatial-mode image. 

Quantum correlated images have been demonstrated using the four-wave mixing process in Ref.~\cite{Boyer2008} and parametric down-conversion processes~\cite{Brida2010, Samantaray2017, Meda2017, Gregory2020}.  The exposure time of the single shot in the imaging system of Ref.~\cite{Samantaray2017} using parametric down-conversion is 100 ms.  The difficulty in that experiment is to achieve a good collection efficiency of the correlated photons and maintain the spatial resolution.  We believe that with the present 4WM scheme a single-frame image with an approximately 1 microsecond long pulse of light onto a CCD camera could provide sub-shot-noise intensity-difference imaging.  In the seeded 4WM process the small phase-matching angle ($\approx0.1^\circ$ to $0.2^\circ$ in potassium vapor~\cite{Swaim2017, Curcic2018}; $\approx0.3^\circ$ to $0.5^\circ$ in rubidium vapor~\cite{Turnbull2013, Kim2018, Wu2019}; $\approx0.3^\circ$ in cesium vapor~\cite{Ma2018}) limits the high spatial frequency parts of the twin images when being reimaged when we use the imaging setup (Fig.~\ref{fig:fig42} (b)).  We can generate twin images of a Gaussian beam using the imaging setup and send one of the twin beams onto an absorbing object or through a transparent mask, and measure the correlations.  

\section{Conclusion}

We have examined dual-seeded twin-beam generation and intensity-difference squeezing in 4WM in a Rb vapor pumped by a Ti:sapphire laser source.  We have identified the noise source from the 2-beam coupling effect that causes excess noise in a dual-seeded 4WM process.  An intensity-dependent 2-beam coupling at low (\hbox{$<$ 5MHz}) frequencies couples bright beams using the same atoms for gain, and would present a problem with any pump system.  The low-frequency 2-beam coupling can be avoided in many circumstances, such as in the case of the dual-seeding for low-frequency squeezing introduced in~\cite{Wu2019}, by seeding skew rays in the gain medium. 

The 2-beam coupling mechanism raises concerns for imaging applications, as mixing the quantum correlations between spatial modes or image pixels could severely impact the possibility of obtaining quantum-enhanced sensitivity or resolution in such systems. 

The 2-beam coupling mechanism should be present in any high nonlinearity 4WM system, not limited to Rb.  While we have pointed out significant noise sources for common implementations of atomic-vapor 4WM, we have also discussed techniques to avoid the added noise in a number of circumstances, and the prospects are still quite promising for using alkali-vapor-based 4WM in real applications, including quantum imaging.

\section*{Funding}
Funding for this work is from the Air Force Office of Scientific Research under grant \#FA9550-16-1-0423.

\section*{Disclosures}
The authors declare that there are no conflicts of interest related to this article.


\bibliography{TwoBeamCoupling}

\end{document}